\documentclass{article}

\usepackage{PRIMEarxiv}

\usepackage[utf8]{inputenc} 
\usepackage[T1]{fontenc}    
\usepackage{hyperref}       
\usepackage{url}            
\usepackage{booktabs}       
\usepackage{amsfonts}       
\usepackage{nicefrac}       
\usepackage{microtype}      
\usepackage{lipsum}
\usepackage{fancyhdr}       
\usepackage{graphicx}       
\graphicspath{{media/}}     
\usepackage{amsmath}
\usepackage{bm}
\usepackage{graphics}
\usepackage{graphicx}
\usepackage{epsfig}
\usepackage{amssymb}
\usepackage{amsthm}
\usepackage{bm}
\usepackage{amsmath}
\usepackage{color}
\usepackage{array}
\usepackage{todonotes}
\usepackage{bm}

\usepackage{caption}

\title{Connecting Weakly Nonlinear Elasticity Theories of Isotropic Hyperelastic Materials
\thanks{\textit{\underline{Citation}}: 
\textbf{Du, Y., Hill, N. A., \& Luo, X. (2024). Connecting weakly nonlinear elasticity theories of isotropic hyperelastic materials. Mathematics and Mechanics of Solids, 29(9), 1900-1914.}} 
}

\author{
  Yangkun Du\thanks{Corresponding author} \\
  DICAM, University of Trento\\
  Trento,  38123, Italy \\
  \texttt{duyangkunzju@gmail.com} 
  \And
  Nicholas A Hill \\
  School of Mathematics and Statistics \\
  University of Glasgow \\
  Glasgow G12 8QQ, UK 
  \And
  Xiaoyu Luo \\
  School of Mathematics and Statistics \\
  University of Glasgow \\
  Glasgow G12 8QQ, UK 
}

\begin{document}
\maketitle

\begin{abstract}
Soft materials exhibit significant nonlinear geometric deformations and stress-strain relationships under external forces. This paper explores weakly nonlinear elasticity theories, including Landau's and Murnaghan's formulations, advancing understanding beyond linear elasticity. We establish connections between these methods and extend strain-energy functions to the third and fourth orders in power of $\varepsilon$, where $\varepsilon=\sqrt{\mathbf{H}\cdot \mathbf{H}}$ and $0<\varepsilon\leq 1$, and $\mathbf{H}$ is the perturbation to the deformation gradient tensor $\mathbf{F} = \mathbf{I} + \mathbf{H}$. Furthermore, we address simplified strain-energy functions applicable to incompressible materials. 
Through this work, we contribute to a comprehensive understanding of nonlinear elasticity and its relationship to weakly nonlinear elasticity, facilitating the study of moderate deformations in soft material behavior and its practical applications.
\end{abstract}

\keywords{Weakly nonlinear elasticity \and Hyperelasticity \and Incompressible materials \and Landau \and Murnaghan \and Invariants}

\section{Introduction}
Soft materials, such as biological tissues and gels, often undergo significant geometric deformations when subjected to external forces \cite{li2012mechanics}. Unlike hard materials, which typically only experience small deformations, the stress-strain relationship in soft materials is best described using nonlinear elasticity theory due to the large deformations involved \cite{kuhl2003theory,alijani2014non,wang2023strain,destrade2023canceling}. Linear elasticity theory is often insufficient to accurately represent the stress-strain relationship in these materials, necessitating the use of nonlinear finite-deformation theory and precise constitutive modeling \cite{ogden1997non}.

For homogeneous isotropic hyperelastic materials, finite-deformation analysis commonly relies on the strain-energy function represented by the three principal invariants of the strain tensor \cite{ogden1997non}. 
The stress-strain relationship can be obtained by solving the partial derivatives of the strain-energy function. 
Under the assumption of incompressibility, the strain-energy can be further simplified as a function of two principal invariants. 
Analytical solutions for large deformation problems of incompressible materials in simple structures have been found based on finite deformation theory \cite{rivlin1953solution}. However, obtaining analytical solutions that account for large deformations becomes difficult for complex problems like nonlinear contact \cite{duan2012effect} and post-buckling analysis \cite{cai1999imperfection}. 
Linear elastic approximations also fail to meet the precision requirements. Instead, weakly nonlinear theory provides an effective approach in such cases \cite{saccomandi2021some}.

In weakly nonlinear elasticity theory, particularly the formulation proposed by \cite{landau1986theory} and \cite{murnaghan1937finite}, represents a significant advance in nonlinear elasticity theory. 
The use of the strain-energy function in polynomial form allows for the precise determination of elastic moduli through curve fitting of experimental data using standard linear regression techniques \cite{saccomandi2021some,ogden2004fitting}.

Landau's approach \cite{landau1986theory}, based on the Landau invariants of the Cauchy-Green strain tensor, with terms up to the third and/or higher orders of the strain-energy functions, includes both material and geometric nonlinearity \cite{saccomandi2021some}. It finds applications in material science, geophysics, acoustics, and other fields, accurately predicting the mechanical response of materials under realistic loading conditions \cite{destrade2002incompressible,destrade2010onset,krishna2012interaction}.

Murnaghan's framework \cite{murnaghan1937finite}, which expresses the strain-energy function as a triply-infinite power series in the principal invariants of the Cauchy-Green strain tensor, is also widely used. This approach has shown the ability to solve simple problems involving compressible materials and specific cross-sectional shapes of prisms under incompressible conditions \cite{rivlin1953solution}. 
It approximates the strain-energy function to any desired order in the power-series expansion, utilizing the symmetric functions of principal invariants. 
Additionally, under the assumption of small deformations, the higher-order Murnaghan model extends the classical linear elastic theory into the weakly nonlinear region, providing a robust method.
By considering the superposition of displacements of higher orders and substituting them into the motion equations and boundary conditions according to the corresponding orders, the Murnaghan model \cite{rivlin1953solution} demonstrates the linearization of the nonlinear problem by neglecting higher-order terms and simplifying solutions to quasi-nonlinear problems \cite{sabin1983contact, du2023nonlinear}.

This article aims to organize the definitions of different strains, invariants, and strain-energy functions in these two different weakly nonlinear elastic theories and the various forms of strain-energy  functions and material parameters in non-linear elastic theory to establish their relationships. 
We expand the strain-energy functions in the weakly nonlinear theory up to the third and fourth orders, corresponding to the second-order and third-order elasticity theories. 
Additionally, we consider the simplified strain-energy functions and stress-strain relationships of materials under incompressible conditions, as many soft materials can be assumed to be incompressible.

\section{Connections among different strain invaraints}

In the context of finite deformation, let us consider an elastic body undergoing a finite displacement field $\boldsymbol{u}$ from the reference configuration to the current configuration. The deformation gradient tensor $\mathbf{F}$ is defined by $\mathbf{F} = \mathbf{I} + \mathbf{H}$, where $\mathbf{H}$ is the displacement gradient tensor, defined as $\mathbf{H} = \mathrm{Grad}, \mathbf{u}$. 
For the purpose of weakly nonlinear theory, we shall assume that $\mathbf{H}$ ia the perturbation to the deformation gradient tensor is small. i.e. $\varepsilon=\sqrt{\mathbf{H}\cdot \mathbf{H}}$ and $0<\varepsilon\leq 1$.
Henceforth, the left and right Cauchy-Green strain tensors are
%
\begin{equation}
\begin{aligned}
\mathbf{b} &=\mathbf{FF^{\text{T}}} =\mathbf{I} +\mathbf{H} +\mathbf{H}^{\text{T}} +\mathbf{HH}^{\text{T}}=\mathbf{I} +\mathbf{e} +\bm{\alpha }, \\
\mathbf{C}&=\mathbf{F^{\text{T}} F} =\mathbf{I} +\mathbf{H} +\mathbf{H}^{\text{T}} +\mathbf{H}^{\text{T}}\mathbf{H}=\mathbf{I} +\mathbf{e} +\bm{\gamma },
\end{aligned}
\label{eq-1}
\end{equation}
where we separate $\mathbf{e} = \mathbf{H} + \mathbf{H}^{\text{T}}$, $\bm{\alpha} = \mathbf{HH}^{\text{T}}$, and $\bm{\gamma }=\mathbf{H}^{\text{T}}\mathbf{H}$, so that $\mathbf{e}$ is the first order term i.e. $\mathbf{e}=O(\varepsilon)$ and $\bm{\alpha}=\bm{\gamma}=O(\varepsilon^2)$ are of second order terms. The Green-Lagrange strain tensor $\mathbf{E}$ is then given by
\begin{equation}
\begin{aligned}
\mathbf{E} =\frac{1}{2}(\mathbf{C} -\mathbf{I}) =\frac{1}{2}\left(\mathbf{e} +\bm{\gamma }\right).
\end{aligned}
\label{eq2}
\end{equation}
In linear elasticity, providing that the displacement gradient tensor $\mathbf{H}$ is small, and ignoring the second-order terms of $\mathbf{H}$, we obtain the infinitesimal strain tensor 
\begin{equation}
\begin{aligned}
\mathbf{E}^{*} =\frac{1}{2}\left( \mathbf{H} +\mathbf{H}^{T}\right)=\frac{1}{2}\mathbf{e}.
\end{aligned}
\end{equation}

The strain-energy function of an ideal isotropic elastic material, capable of undergoing finite deformation, can generally be expressed in terms of three independent scalar invariants of the strain tensor. For a general second-order tensor $\mathbf{M}$, the principal invariants are defined by the equation
\begin{equation}
\mathbf{M}^{3} -I_{M}\mathbf{M}^{2} +II_{M}\mathbf{M} -III_{M}\mathbf{I} = \mathbf {0},
\label{eq4}
\end{equation}
where the invariants of the tensor  $\mathbf{M}$ are 
\begin{equation}
\begin{aligned}
I_{M} =\mathrm{tr}(\mathbf{M}), \quad II_{M} =\frac{1}{2}\left(\mathrm{tr}(\mathbf{M})\mathrm{^{2} -tr}\left(\mathbf{M}^{2}\right)\right),\quad III_{M} =\mathrm{det}\mathbf{M}.
\end{aligned}
\end{equation}
On the other hand, the scalar invariants of the tensor  $\mathbf{M}$ can be defined in Landau's form by
\begin{equation}
\begin{aligned}
\bar I_{1} =\mathrm{tr}(\mathbf{M}),\quad \bar I_{2} =\mathrm{tr}\left(\mathbf{M}^{2}\right),\quad \bar I_{3} =\mathrm{tr}\left(\mathbf{M}^{3}\right),
\end{aligned}
\end{equation}
where $\bar I_1$, $\bar I_2$, and $\bar I_3$ are respectively the first-, second- and third-order terms of $\mathbf{M}$.

Some commonly used invariants and the relationships between them are
\begin{equation}
\begin{aligned}
&I_C=\mathrm{tr}(\mathbf{C}),\quad II_C=\frac{1}{2}\left( \mathrm{tr(}\mathbf{C}\mathrm{)^{2} -tr}\left(\mathbf{C}^{2}\right)\right),\quad III_C=\mathrm{det}\mathbf{C},\\
&\bar{I}_{1} =\mathrm{tr}(\mathbf{E}) ,\quad \bar{I}_{2} =\mathrm{tr}\left(\mathbf{E}^{2}\right) ,\quad \bar{I}_{3} =\mathrm{tr}\left(\mathbf{E}^{3}\right),\\
&J_{1} =\mathrm{tr}( 2\mathbf{E}) ,\quad J_{2} =\frac{1}{2}\left(\mathrm{ tr(} 2\mathbf{E}\mathrm{)^{2} -tr}\left( 4\mathbf{E}^{2}\right)\right) ,\quad J_{3} =\mathrm{det}( 2\mathbf{E}),\\
&\tilde{J}_{1} =\mathrm{tr}(\mathbf{E}) ,\quad\tilde{J}_{2} =\frac{1}{2}\left(\mathrm{ tr(}\mathbf{E}\mathrm{)^{2} -tr}\left(\mathbf{E}^{2}\right)\right) ,\quad\tilde{J}_{3} =\mathrm{det}\mathbf{E}.
\end{aligned}
\label{eq6}
\end{equation}
Among them, $I_C, II_C$, and $III_C$ are principal invariants of the right Cauchy-Green strain tensor $\mathbf{C}$, which are often used for constructing the strain-energy function for hyperelastic material in fully nonlinear elasticity. 
When the material is incompressible ($III_C=1$), the strain-energy function can be simplified to depend solely on $I_C$ and $II_C$.
Additionally, the Landau invariants $\bar{I}_1, \bar{I}_2$, and $\bar{I}_3$ represent the first, second and third-order terms of the Green-Lagrange strain tensor $\mathbf{E}$.
Similarly, the principal invariants of the Green-Lagrange strain tensor $\mathbf{E}$, namely $\tilde{J}_1, \tilde{J}_2$, and $\tilde{J}_3$, also represent its first, second and third-order terms. These are known as Murnaghan invariants.
However, it is more convenient to use another set of Murnaghan invariants, namely ${J}_1, {J}_2$, and ${J}_3$, which are the principal invariants of $2\mathbf{E}$ to simplify the derivation of the stress-strain relationships. 

Given the often utilization of both fully nonlinear and weakly nonlinear elasticity theories in the derivation of nonlinear deformation problems, it becomes imperative to comprehensively outline the interconnections between the invariants and the strain-energy functions they engender.

\subsection{Connections between $I_C,~ II_C$, $III_C$ and $\bar{I}_1, ~\bar{I}_2$, $\bar{I}_3$}
In this subsection, we shall demonstrate the connections between the principal invariants, $I_C$, $II_C$, and $III_C$, of the right Cauchy-Green strain tensor $\mathbf{C}$ and the Landau invariants, $\bar{I}_1$, $\bar{I}_2$, $\bar{I}_3$, of the Green-Lagrange strain tensor $\mathbf{E}$.
First, recalling the Eqs. \eqref{eq2} and \eqref{eq6}, we have 
\begin{equation}
\begin{aligned}
\mathrm{tr}\mathbf{C} &=\mathrm{tr}(\mathbf{I} +2\mathbf{E}) =2\bar{I}_{1} +3,\\
\mathrm{tr}\left(\mathbf{C}^{2}\right) &=\mathrm{tr}\left( 4\mathbf{E}^{2} +4\mathbf{E} +\mathbf{I}\right) =4\bar{I}_{2} +4\bar{I}_{1} +3,\\
\mathrm{tr}\left(\mathbf{C}^{3}\right) &=\mathrm{tr}\left( 8\mathbf{E}^{3} +12\mathbf{E}^{2} +6\mathbf{E} +\mathbf{I}\right) =8\bar{I}_{3} +12\bar{I}_{2} +6\bar{I}_{1} +3.
\end{aligned}
\end{equation}
Then, the first and second principal invariants of the right Cauchy-Green strain tensor $\mathbf{C}$ can be expressed as 
\begin{equation}
\begin{aligned}
I_{C} &=\mathrm{tr}\mathbf{C} =2\bar{I}_{1}+3,\\
II_{C} &=\frac{1}{2}\left( \mathrm{tr(}\mathbf{C}\mathrm{)^{2} -tr}\left(\mathbf{C}^{2}\right)\right) =3+4\bar{I}_{1} +2\bar{I}_{1}^{2} -2\bar{I}_{2}.\\
\end{aligned}
\label{IC-IIC-Ib}
\end{equation}
Next, by tracing the Eq. \eqref{eq4}, we obtain 
\begin{equation}
\mathrm{tr}\left(\mathbf{C}^{3}\right) -I_{C}\mathrm{tr}\left(\mathbf{C}^{2}\right) +II_{C}\mathrm{tr}\mathbf{C} -3III_{C} =0.
\end{equation}
Using this equation, the third principal invariant of the right Cauchy-Green strain tensor $\mathbf{C}$ can be expressed as
\begin{equation}
\begin{aligned}
III_{C} &=\mathrm{det}\mathbf{C} =\frac{1}{3}\left(\mathrm{tr}\left(\mathbf{C}^{3}\right) -I_{C}\mathrm{tr}\left(\mathbf{C}^{2}\right) +II\mathrm{_{C} tr}\mathbf{C}\right)\\
&=1+2\bar{I}_{1} +2\bar{I}_{1}^{2} -2\bar{I}_{2} +\frac{4}{3}\bar{I}_{1}^{3} -4\bar{I}_{1}\bar{I}_{2} +\frac{8}{3}\bar{I}_{3}.
\end{aligned}
\label{IIIC-Ib}
\end{equation}
Inversely, from to Eqs. \eqref{IC-IIC-Ib} and \eqref{IIIC-Ib}, we can express the Landau invariants $\bar{I}_1, ~\bar{I}_2$, and $\bar{I}_3$ in terms of  the principal invariants $I_C,~ II_C$, and $III_C$ as
\begin{equation}
\begin{aligned}
\bar{I}_{1} &=\frac{1}{2}( -3+I_{C}),~ 
\bar{I}_{2} =\frac{1}{4}\left( 3-2I_{C} +I_{C}^{2} -2II_{C}\right),\\
\bar{I}_{3} &=\frac{1}{8}\left( 24-24I_{C} +12I_{C}^{2} -2I_{C}^{3}
-12II_{C} +3I_{C} II_{C} +3III_{C}\right).
\end{aligned}
\label{Ib-IC-IIC-IIIC}
\end{equation}

\subsection{Connections between $I_C, II_C$, $III_C$, $J_1, J_2$, $J_3$, and  $\tilde J_1, \tilde J_2$, $\tilde J_3$}

As the two sets of definitions of the Murnaghan invariants $J_1, J_2$, $J_3$, and  $\tilde J_1, \tilde J_2$, $\tilde J_3$ in Eq. \eqref{eq6} are commonly used, we shall first show connections between them and then demonstrate relationships with to the principal invariants of the right Cauchy-Green strain tensor $\mathbf{C}$.

From Eq. \eqref{eq6}, we obtain 
\begin{equation}
\begin{aligned}
J_{1} &=\mathrm{tr}( 2\mathbf{E})=2\mathrm{tr}\mathbf{E}=2\tilde{J}_{1},\\
J_{2} &=\frac{1}{2}\left( \mathrm{tr} (2\mathbf{E})^{2} -\mathrm{tr}\left( 4\mathbf{E}^{2}\right)\right)=\frac{1}{2}\left(4\tilde{J}_{1}^{2} -4\mathrm{tr}\left(\mathbf{E}^{2}\right)\right) =4\tilde{J}_{2},\\
J_{3} &=\mathrm{det}( 2\mathbf{E}) =8\mathrm{det}\mathbf{E} =8\tilde{J}_{3}.
\end{aligned}
\label{J-tJ}
\end{equation}
Furthermore, utilizing Eqs. \eqref{eq2} and \eqref{eq6}, we can deduce the following relationships between the Murnaghan invariants $I_C, II_C$, $III_C$ and $J_{1}, J_{2}, J_{3}$:
\begin{equation}
\begin{aligned}
I_{C} &=\mathrm{tr}(\mathbf{I} +2\mathbf{E}) =3+\mathrm{tr}( 2\mathbf{E}) =3+J_{1},\\
II_{C} & =\frac{1}{2}\left(( J_{1} +3)^{2} -\mathrm{tr}\left( 4\mathbf{E}^{2} +4\mathbf{E} +\mathbf{I}\right)\right)=3+2J_{1}+J_{2},\\
III_{C} &=\mathrm{det}(\mathbf{I} +2\mathbf{E}) =1+J_{1} +J_{2} +J_{3}.
\end{aligned}
\label{I-J}
\end{equation}
Inversely, we have 
\begin{equation}
\begin{aligned}
J_{1} &=I_{C} -3,~ 
J_{2} &=3-2I_{C} +II_{C},~ 
J_{3} &=I_{C} -II_{C} +III_{C} -1.
\end{aligned}
\end{equation}
Moreover, considering the connections between $J_i$ and $\tilde J_i(i=1,2,3)$ in Eq. \eqref{J-tJ}, we have 
\begin{equation}
\begin{aligned}
I_{C} &=3+2\tilde{J}_{1},~ 
II_{C} &=3+4\tilde{J}_{1} +4\tilde{J}_{2},~ 
III_{C} &=1+2\tilde{J}_{1} +4\tilde{J}_{2} +8\tilde{J}_{3},
\end{aligned}
\end{equation}
and 
\begin{equation}
\begin{aligned}
\tilde{J}_{1} &=\frac{1}{2}( I_{C} -3),~ 
\tilde{J}_{2} &=\frac{1}{4}( 3-2I_{C} +II_{C}),~ 
\tilde{J}_{3} &=\frac{1}{8}( I_{C} -II_{C} +III_{C} -1).
\end{aligned}
\label{J-I}
\end{equation}


\subsection{Connections among $\bar{I}_{1},~ \bar{I}_{2}$, $\bar{I}_{3}$, $J_1,~ J_2$, $J_3$, and  $\tilde J_1,~ \tilde J_2$, $\tilde J_3$}
This subsection explores the connections between the Landau invariants $\bar{I}_{1},~ \bar{I}_{2}$, $\bar{I}_{3}$ and the Murnaghan invariants $J_1,~ J_2$, $J_3$. Utilizing Eqs. \eqref{IC-IIC-Ib}, \eqref{IIIC-Ib}, and \eqref{I-J}, the following relationships can be established:
\begin{equation}
\begin{aligned}
2\bar{I}_{1}+3 &=3+J_{1},\\
3+4\bar{I}_{1} +2\bar{I}_{1}^{2} -2\bar{I}_{2} &=3+2J_{1}+J_{2},\\
1+2\bar{I}_{1} +2\bar{I}_{1}^{2} -2\bar{I}_{2} +\frac{4}{3}\bar{I}_{1}^{3} -4\bar{I}_{1}\bar{I}_{2} +\frac{8}{3}\bar{I}_{3} &=1+J_{1} +J_{2} +J_{3}.
\end{aligned}
\end{equation}
Solving these equations yields
\begin{equation}
\begin{aligned}
\bar{I}_{1} &=\frac{1}{2} J_{1},~ 
\bar{I}_{2} &=\frac{1}{4}\left( J_{1}^{2} -2J_{2}\right), ~
\bar{I}_{3} &=\frac{1}{8}\left( J_{1}^{3} -3J_{1} J_{2} +3J_{3}\right),
\end{aligned}
\label{Ib-J}
\end{equation}
and 
\begin{equation}
\begin{aligned}
J_{1} &=2\bar{I}_{1},~ 
J_{2} &=2\left(\bar{I}_{1}^{2} -\bar{I}_{2}\right),~ 
J_{3} &=\frac{4}{3}\left(\bar{I}_{1}^{3} -3\bar{I}_{1}\bar{I}_{2} +2\bar{I}_{3}\right).
\end{aligned}
\label{J-Ib}
\end{equation}
Analogously, according to the Eq. \eqref{J-tJ}, we can obtain the connections between the Landau invariants $\bar I_i(i=1,2,3)$ and Murnaghan invariants $\tilde J_i(i=1,2,3)$ are
\begin{equation}
\begin{aligned}
\bar{I}_{1} &=\tilde{J}_{1},~ 
\bar{I}_{2} &=\tilde{J}_{1}^{2} -2\tilde{J}_{2},~ 
\bar{I}_{3} &=\tilde{J}_{1}^{3} -3\tilde{J}_{1}\tilde{J}_{2} +3\tilde{J}_{3}.
\end{aligned}
\end{equation}
and
\begin{equation}
\begin{aligned}
\tilde{J}_{1} &=\bar{I}_{1},~ 
\tilde{J}_{2} &=\frac{1}{2}\left(\bar{I}_{1}^{2} -\bar{I}_{2}\right),~ 
\tilde{J}_{3} &=\frac{1}{6}\left(\bar{I}_{1}^{3} +3\bar{I}_{1}\bar{I}_{2} +2\bar{I}_{3}\right).
\end{aligned}
\end{equation}

Finally, the summary of transformations for the principal invariants, $I_C$, $II_C$, and $III_C$, of the right Cauchy-Green strain tensor $\mathbf{C}$, the Landau invariants $\bar{I}_1$, $\bar{I}_2$, $\bar{I}_3$ of the Green-Lagrange strain tensor $\mathbf{E}$, and the Murnaghan invariants $J_1, J_2, J_3$ of the strain tensor $2 \mathbf{E}$ can be found in Table \ref{table-1}.

\normalsize
\section{Weakly nonlinear elasticity for isotropic compressible materials}
In fully nonlinear elasticity, the strain-energy function of the isotropic compressible material $W$ is commonly expressed in terms of the three principal invariants ($I_C$, $II_C$, and $III_C$) of the right Cauchy-Green strain tensor $\mathbf{C}$. The Cauchy stress tensor $\mathbf{t}$ can be represented as
\begin{equation}
 \mathbf{t} =\frac{2}{\mathcal{J}}\left(\mathbf{b}\frac{\partial W}{\partial I_{C}} -III_{C}\mathbf{b}^{-1}\frac{\partial W}{\partial II_{C}} +\left( III_{C}\frac{\partial W}{\partial III_{C}} +II_{C}\frac{\partial W}{\partial II_{C}}\right)\mathbf{I}\right),
 \label{CauchyS}
\end{equation}
where $\mathbf{b}$ is the left Cauchy-Green tensor. $\mathcal{J}=\mathrm{det} \mathbf{F}$ and $\mathcal{J}=1$ for incompressible materials. 
Here $ I_{C}-3$, $II_{C}-3$, and $III_{C}-1$ are all of $O(\varepsilon)$. 
For weakly nonlinear elasticity, instead of using the principal invariants of $\mathbf{C}$ to construct the strain-energy function and stress tensor, it is more convenient to employ the Landau and Murnaghan invariants for constructing the strain-energy function up to a certain order of expansion.

\subsection{Second-order elasticity}
\subsubsection{strain-energy functions}
In second-order elasticity, we require the terms of  $O(\varepsilon^2)$ in the stress and strain tensors. Consequently, the strain-energy function needs to include the third-order terms of $O(\varepsilon^3)$. In terms of Murnaghan invariants, the strain-energy function to third-order terms is
\begin{equation}
W_{M} =a_{1} J_{2} +a_{2} J_{1}^{2} +a_{3} J_{1} J_{2} +a_{4} J_{1}^{3} +a_{5} J_{3}+O(\varepsilon^4)
\label{Murn-3}
\end{equation}
where $a_1, ..., a_5$ are $O(\varepsilon)$ material constants. 
$W_{2M} =a_{1} J_{2} +a_{2} J_{1}^{2}=O(\varepsilon^2)$ and $W_{3M} =a_{3} J_{1} J_{2} +a_{4} J_{1}^{3} +a_{5} J_{3}=O(\varepsilon^3)$ are the second and the third order terms, respectively.
%
Alternatively, third-order energy function can be expressed in terms of Landau invariants as:
\begin{equation}
W_{L} =\frac{\lambda }{2}\bar{I}_{1}^{2} +\mu \bar{I}_{2} +\frac{\bar{A}}{3}\bar{I}_{3} +\bar{B}\bar{I}_{1}\bar{I}_{2} +\frac{\bar{C}}{3}\bar{I}_{1}^{3}+O(\varepsilon^4),
\end{equation}
where $\mu$ and $\lambda$ are the linear Lam\'{e} coefficients, and $\bar A,~ \bar B$ and $\bar C$ are $O(\varepsilon)$ elastic material constants. 
$W_{2L} =\frac{\lambda }{2}\bar{I}_{1}^{2} +\mu \bar{I}_{2}$ is the second order term and 
$W_{3L} =\frac{\bar{A}}{3}\bar{I}_{3} +\bar{B}\bar{I}_{1}\bar{I}_{2} +\frac{\bar{C}}{3}\bar{I}_{1}^{3}$ is the third order term.

Referring to the connections between Murnaghan invariants and Landau invariants in Eq. \eqref{J-Ib}, the third-order strain-energy function in Murnaghan expansion can be rewritten as
\begin{equation}
W_{M} =( 2a_{1} +4a_{2})\bar{I}_{1}^{2} -2a_{1}\bar{I}_{2} +\frac{8a_{5}}{3}\bar{I}_{3} -4( a_{3} +a_{5})\bar{I}_{1}\bar{I}_{2} +\left( 4a_{3} +8a_{4} +\frac{4a_{5}}{3}\right)\bar{I}_{1}^{3}+O(\varepsilon^4).
\end{equation}
Thus, we can obtain the following relationships among the material constants
\begin{equation}
\begin{aligned}
\lambda &=4a_{1} +8a_{2} ,~ \mu =-2a_{1} ,\\
\bar{A} &=8a_{5} ,~ \bar{B} =-4( a_{3} +a_{5}),~\bar{C} =( 12a_{3} +24a_{4} +4a_{5}).
\end{aligned}
\end{equation}
Similarly, referring to the connections between Landau invariants and Murnaghan invariants in Eq. \eqref{Ib-J}, we can rewrite the third-order strain-energy function in Landau expansion as:
\begin{equation}
W_{L} =-\frac{\mu }{2} J_{2} +\left(\frac{\lambda }{8} +\frac{\mu }{4}\right) J_{1}^{2} -\left(\frac{\bar{A}}{8} +\frac{\bar{B}}{4}\right) J_{1} J_{2} +\left(\frac{\bar{A}}{24} +\frac{\bar{B}}{8} +\frac{\bar{C}}{24}\right) J_{1}^{3} +\frac{\bar{A}}{8} J_{3}+O(\varepsilon^4).
\end{equation}
This leads to
\begin{equation}
\begin{aligned}
a_{1} &=-\frac{\mu }{2} ,~\ a_{2} =\frac{\lambda }{8} +\frac{\mu }{4} ,\\
a_{3} &=-\left(\frac{\bar{A}}{8} +\frac{\bar{B}}{4}\right) ,~ a_{4} =\frac{\bar{A}}{24} +\frac{\bar{B}}{8} +\frac{\bar{C}}{24},~ a_{5} =\frac{\bar{A}}{8}.
\end{aligned}
\label{eq-30}
\end{equation}

\subsubsection{Stress-strain relationships}
Second-order elasticity requires the stress and strain tensors to be expanded to  $O(\varepsilon^2)$. According to Eq. \eqref{CauchyS}, the Cauchy stress tensor involves the three principal invariants of the right Cauchy-Green strain tensor ($I_C$, $II_C$, and $III_C$), as well as the partial derivatives of the strain-energy function $W$ with respect to them, the left Cauchy-Green strain tensor $\mathbf{b}$, and its inverse tensor $\mathbf{b}^{-1}$.
For isotropic elastic solids, the principal invariants of the left Cauchy-Green strain tensor $\mathbf{b}$ equal to those of the right Cauchy-Green strain tensor $\mathbf{C}$, i.e., $I_{b} = I_{C}$, $II_{b} = II_{C}$, and $III_{b} =III_{C}$.
From \eqref{eq-1} and defining  
\begin{equation} \begin{aligned}
e=\mathrm{tr}\mathbf{e} ,~\alpha =\mathrm{tr}\bm{\alpha } ,~ \mathbf{K} =(\det\mathbf{e})\mathbf{e}^{-1} ,
\label{eq-34}
\end{aligned} \end{equation} 
the first principal invariant $I_b$ can be expressed as 
\begin{equation} \begin{aligned}
I_{b} =\mathrm{tr}\mathbf{b} =3+e+\alpha,
\label{I_b}
\end{aligned} \end{equation} 
and 
\begin{equation} \begin{aligned}
\mathbf{b}^{2} =(\mathbf{I+e}+\bm\alpha)^{2} =\mathbf{I}+2 \mathbf e+2\bm \alpha  +\mathbf{e}^{2} +O\left( \varepsilon^{3}\right).
\label{b^2}
\end{aligned} \end{equation} 
By tracing this equation and using the Eq. \eqref{eq-34}, we obtain
 \begin{equation} \begin{aligned}
\mathrm{tr}\left(\mathbf{b}^{2}\right) =3+2e+\alpha +e^{2} -2K+O\left( \varepsilon^{3}\right),
 \end{aligned} \end{equation} 
where $K=\mathrm{tr}\mathbf{K} =\frac{1}{2}\left((\mathrm{tr}\mathbf{e})^{2} -\mathrm{tr}\left(\mathbf{e}^{2}\right)\right)$. Thus, the second principal invariant $II_b$ is given as 
\begin{equation} \begin{aligned}
II_{b} =\frac{1}{2}\left( (\mathrm{tr}\mathbf{b})^{2} -\mathrm{tr}\left(\mathbf{b}^{2}\right)\right)  =3 +2e+2\alpha +K+O(\varepsilon^3),
\label{II_b}
\end{aligned} \end{equation} 
and the third principal invariant $III_b$ is  
\begin{equation} \begin{aligned}
III_{b} =\mathrm{det}\mathbf{b} =\mathrm{det}(\mathbf{I+e+\bm\alpha }) =1+j_{1} +j_{2} +j_{3},
\label{eq-37}
\end{aligned} \end{equation}
where $j_{1} =\mathrm{tr}(\mathbf{e+\bm\alpha }) ,\ j_{2} =\frac{1}{2}\left( j_{1}^{2} -\mathrm{tr}\left((\mathbf{e+\bm\alpha })^{2}\right)\right) ,\ j_{3} =\mathrm{det}(\mathbf{e+\bm\alpha })$.
Referring to Eq. \eqref{eq-34} and expanding to second-order terms, we have
\begin{equation} \begin{aligned}
j_{1} =e+\alpha+O\left( \varepsilon^{3}\right),~ j_{2}= K+O\left( \varepsilon^{3}\right),~ j_{3}=0+O\left( \varepsilon^{3}\right).
\end{aligned} \end{equation}
Thus, the third principal invariant $III_b$ can be expanded as
\begin{equation} \begin{aligned}
III_{b} =1+e+\alpha+K+O\left( \varepsilon^{3}\right)
\label{eq-39}
\end{aligned} \end{equation}
In addition, using the Cayley-Hamilton theorem and Eqs. \eqref{I_b}, \eqref{b^2}, and \eqref{II_b}, we have
\begin{equation}
\begin{aligned}
III_{C}\mathbf{b}^{-1} &=III_{b}\mathbf{b}^{-1}=\mathbf{b}^{2} -I_{b}\mathbf{b} +II_{b}\mathbf{I}\\
&=(\mathbf{e} -e-1)\mathbf{e} -\bm{\alpha } +( 1+ e+\alpha +K)\mathbf{I} +O\left( \varepsilon^{3}\right).
\label{eq-40}
\end{aligned}
\end{equation}
Analogously, from Eq.\eqref{eq-34}, we have 
\begin{equation} \begin{aligned}
\mathbf{K}=(\mathrm{det}\mathbf{e})\mathbf{e}^{-1}=\mathbf{e}^{2} -\mathnormal{I_{e}}\mathbf{e} +\mathnormal{II_{e}}\mathbf{I},
\end{aligned} \end{equation} 
where $
I_{e} =\mathrm{tr}\mathbf{e} =e,~
II_{e} =\frac{1}{2}\left((\mathrm{tr}\mathbf{e})^{2} -\mathrm{tr}\left(\mathbf{e}^{2}\right)\right) =K$.
This equation gives
\begin{equation} \begin{aligned}
\mathbf{e}^{2}=\mathbf{K}+e\mathbf{e}-K\mathbf{I}.
\end{aligned} \end{equation} 
Hence,  Eq. \eqref{eq-40} can be finally rewritten as
\begin{equation} \begin{aligned}
\mathnormal{III_{C}}\mathbf{b}^{-1} =\mathbf{K} -\mathbf{e} -\mathbf{\bm\alpha } +( 1+ e+\alpha )\mathbf{I} +O\left( \varepsilon^{3}\right) .
\label{eq-42}
\end{aligned} \end{equation} 

Next, following the connections between the Murnaghan invariants and the principal invariants in Eq. \eqref{J-I}, we have 
\begin{equation}
 \begin{aligned}
J_{1} &=I_{C} -3=e+\alpha, \\
J_{2} &=3-2I_{C} +II_{C} =K+O\left( \varepsilon^{3}\right),\\
J_{3} &=I_{C} -II_{C} +III_{C} -1=O\left( \varepsilon^{3}\right).
\end{aligned}
\end{equation}
In addition, using the chain rule, we can replace the derivatives of the strain-energy function $W$ with respect to  $I_C$, $II_C$, and $III_C$ by 
\begin{equation}
 \begin{aligned}
\frac{\partial W_{M}}{\partial I_{C}} &=\frac{\partial W_{M}}{\partial J_{1}} -2\frac{\partial W_{M}}{\partial J_{2}} +\frac{\partial W_{M}}{\partial J_{3}},\\
\frac{\partial W_{M}}{\partial II_{C}} &=\frac{\partial W_{M}}{\partial J_{2}} -\frac{\partial W_{M}}{\partial J_{3}},~ 
\frac{\partial W_{M}}{\partial III_{C}} =\frac{\partial W_{M}}{\partial J_{3}}.
\end{aligned}
\end{equation}
With respect to the third-order Murnaghan strain-energy function in Eq. \eqref{Murn-3}, we have 
\begin{equation}
\begin{aligned}
\frac{\partial W_{M}}{\partial I_{C}} &=( a_{5} -2a_{1}) +2( a_{2} -a_{3}) J_{1} +a_{3} J_{2} +3a_{4} J_{1}^{2}+O\left( \varepsilon^{3}\right),\\
 \frac{\partial W_{M}}{\partial II_{C}}&=(a_{1}-a_{5})+a_{3}J_{1}+O\left( \varepsilon^{2}\right),~
\frac{\partial W_{M}}{\partial III_{C}}=a_{5}+O\left( \varepsilon\right).
\end{aligned}
\label{dW/dJ}
\end{equation}
Moreover, recalling that $\mathcal{J}\mathrm{=det}(\mathbf{F}) =III_{C}^{1/2}$ and using \eqref{eq-39}, we have
\begin{equation} 
\begin{aligned}
\frac{2}{\mathcal{J}} =\frac{2}{\mathrm{det}(\mathbf{F})} =2-e+\frac{3}{4} e^{2} -K-\alpha +O\left( \varepsilon^{3}\right).
\label{2/J}
\end{aligned} 
\end{equation}
Finally, substituting Eqs. \eqref{I_b}, \eqref{II_b}, \eqref{eq-39}, \eqref{eq-42}, \eqref{dW/dJ}, and \eqref{2/J} into \eqref{CauchyS} and ignoring the higher-order terms, we can derive the second-order expansion of the Cauchy stress tensor 
\begin{equation}
\begin{aligned}
\mathbf{t} =&\left[ -2a_{1}\mathbf{e} +( 2a_{1} +4a_{2}) e\mathbf{I}\right] +\left[( a_{1} +4a_{2} -2a_{3}) e\mathbf{e} -( 2a_{1} -2a_{5})\mathbf{K} -2a_{1}\bm{\alpha } \right.\\
&\left.+\left(( 2a_{1} +2a_{3}) K+( 2a_{1} +4a_{2}) \alpha -( a_{1} +2a_{2} -2a_{3} -6a_{4}) e^{2}\right)\mathbf{I}\right]+O\left( \varepsilon^{3}\right).
\end{aligned}
\end{equation}
From this, we  extract the first-order linear elastic term 
\begin{equation}
\begin{aligned}
\mathbf{t}_{1} =& -2a_{1}\mathbf{e} +( 2a_{1} +4a_{2}) e\mathbf{I},
\end{aligned}
\end{equation}
and the second-order term
\begin{equation}
\begin{aligned}
\mathbf{t} _{2}=&( a_{1} +4a_{2} -2a_{3}) e\mathbf{e} -( 2a_{1} -2a_{5})\mathbf{K} -2a_{1}\bm{\alpha }\\
&+\left(( 2a_{1} +2a_{3}) K+( 2a_{1} +4a_{2}) \alpha -( a_{1} +2a_{2} -2a_{3} -6a_{4}) e^{2}\right)\mathbf{I}.
\end{aligned}
\end{equation}

\subsection{Third-order elasticity}
\subsubsection{strain-energy functions}
The third-order elasticity requires expanding the stress and strain tensors up to third-order smallness $O\left( \varepsilon^{3}\right)$, as well as the strain-energy function up to fourth-order smallness $O\left( \varepsilon^{4}\right)$. The fourth-order energy function can be expressed using the Murnaghan invariants by
\begin{equation}
W_{M} =W_{2M} +W_{3M}+W_{4M}+O\left( \varepsilon^{5}\right),
\label{Murn-4}
\end{equation}
where $W_{4M}=a_{6} J_{1} J_{3} +a_{7} J_{1}^{2} J_{2} +a_{8} J_{2}^{2} +a_{9} J_{1}^{4}$ and $a_{6},~a_{7},~a_{8}$, and $a_{9}$ are $O\left( \varepsilon\right)$ material constants.
Additionally, the fourth-order energy function can be expressed in terms of the Landau invariants as
\begin{equation}
W_{L} =W_{2L} +W_{3L}+W_{4L}+O\left( \varepsilon^{5}\right),
\end{equation}
%
where $W_{4L} =\bar{E}\bar{I}_{1}\bar{I}_{3} +\bar{F}\bar{I}_{1}^{2}\bar{I}_{2} +\bar{G}\bar{I}_{2}^{2} +\bar{H}\bar{I}_{1}^{4}$ $\mu$ and $\bar E,~\bar F,~\bar G$ and $\bar H$ are $O\left( \varepsilon\right)$ material constants.
%
Referring to the connections between the Murnaghan invariants and Landau invariants in Eq. \eqref{J-Ib}, we can rewrite the fourth-order strain-energy term in the Murnaghan expansion as 
\begin{equation}
\begin{aligned}
W_{4M} &=\frac{16a_{6}}{3}\bar{I}_{1}\bar{I}_{3} -8( a_{6} +a_{7} +a_{8})\bar{I}_{1}^{2}\bar{I}_{2} +4a_{8}\bar{I}_{2}^{2} +\left(\frac{8a_{6}}{3} +8a_{7} +4a_{8} +16a_{9}\right)\bar{I}_{1}^{4}.
\end{aligned}
\end{equation}
Thus,  the relationships between the material constants are
\begin{equation}
\begin{aligned}
\bar{E} \ &=\frac{16a_{6}}{3} ,\ \bar{F} =-8( a_{6} +a_{7} +a_{8}) ,\ \bar{G} =4a_{8} ,\ \bar{H} =\frac{8a_{6}}{3} +8a_{7} +4a_{8} +16a_{9}.
\end{aligned}
\end{equation}
Similarly, referring to the connections between the Landau invariants and Murnaghan invariants in Eq. \eqref{Ib-J}, we can rewrite the fourth-order strain-energy term in the Landau expansion as follows:
\begin{equation}
\begin{aligned}
W_{4L} &=\frac{3\bar{E}}{16} J_{1} J_{3} -\frac{1}{16}( 3\bar{E} +2\bar{F} +4\bar{G}) J_{1}^{2} J_{2} +\frac{\bar{G}}{4} J_{2}^{2} +\frac{1}{16}(\bar{E} +\bar{F} +\bar{G} +\bar{H}) J_{1}^{4},
\end{aligned}
\end{equation}
which gives the relationship among the material constants
\begin{equation}
\begin{aligned}
a_{6} &=\frac{3\bar{E}}{16} ,\ a_{7} =-\frac{1}{16}( 3\bar{E} +2\bar{F} +4\bar{G}) ,\ a_{8} =\frac{\bar{G}}{4} ,\ a_{9} =\frac{1}{16}(\bar{E} +\bar{F} +\bar{G} +\bar{H}).
\end{aligned}
\end{equation}

\subsubsection{Stress-strain relationships}
Here, our aim is to expand all terms in Equation \eqref{CauchyS} up to the third-order terms and obtain the third-order expansion of the Cauchy stress tensor, represented by $\mathbf{t}=\mathbf{t} _1+\mathbf{t} _2+\mathbf{t}_3+O\left( \varepsilon^{4}\right)$, where $\mathbf{t}_3$ denotes the third-order term.
By expanding the square of the left Cauchy-Green strain tensor $\mathbf{b}$, we have 
\begin{equation}
\begin{aligned}
\mathbf{b}^{2}=\mathbf{I}+2\mathbf {e}+2\bm\alpha +\mathbf{e}^{2} +2\mathbf{\bm\beta } +O\left( \varepsilon^{4}\right),
\end{aligned}
\end{equation}
and its trace
 \begin{equation} \begin{aligned}
\mathrm{tr}\left(\mathbf{b}^{2}\right) =3+2e+2\alpha+e^{2}-2K+2\beta+O\left( \varepsilon^{4}\right),
 \end{aligned} \end{equation} 
 where $\bm \beta=\mathbf e \bm \alpha$ and $\beta=\mathrm{tr} (\mathbf e \bm \alpha)$.
Therefore, the second principal invariant $II_b$ can be obtained by
\begin{equation} \begin{aligned}
II_{b} =\frac{1}{2}\left( (\mathrm{tr}\mathbf{b})^{2} -\mathrm{tr}\left(\mathbf{b}^{2}\right)\right)  =3 +2e+2\alpha +K+\alpha e-\beta+O\left( \varepsilon^{4}\right).
\label{II_b-3}
\end{aligned} \end{equation} 
Next, referring to Eqs. \eqref{eq-34} and \eqref{eq-37} and expand to the third-order terms, we find
\begin{equation} \begin{aligned}
j_{1} =e+\alpha,~ j_{2}=K+e\alpha -\beta +O\left( \varepsilon^{4}\right),~ j_{3}=L+O\left( \varepsilon^{4}\right),
\end{aligned} \end{equation}
where $L=\mathrm{det} \mathbf{e}$. Therefore, we can yield 
\begin{equation} \begin{aligned}
III_{b} =1+e+\alpha+K+e\alpha -\beta+L+O\left( \varepsilon^{4}\right).
\label{III_b-3}
\end{aligned} \end{equation}
Moreover, according to the Cayley-Hamilton theorem, we have
\begin{equation}
\begin{aligned}
III_{C}\mathbf{b}^{-1} =&\mathbf{K} -\mathbf{e} -\bm{\alpha } +( 1+ e+\alpha +\alpha e-\beta )\mathbf{I} \\
&+2\mathbf{\bm\beta } -e\mathbf{\bm\alpha } -\alpha \mathbf{e} +O\left( \varepsilon^{4}\right).
\end{aligned}
\label{eq-59}
\end{equation}

By following the connections between the Murnaghan invariants and principal invariants in Eq. \eqref{J-I} and expanding to third-order terms, we find
\begin{equation}
 \begin{aligned}
J_{1} &=I_{C} -3=e+\alpha \\
J_{2} &=3-2I_{C} +II_{C} =K+\alpha e-\beta +O\left( \varepsilon^{4}\right)\\
J_{3} &=I_{C} -II_{C} +III_{C} -1=L+O\left( \varepsilon^{4}\right)
\end{aligned}
\label{eq-60}
\end{equation}
Additionally, using the chain rule, we can replace the derivatives of the strain-energy function $W_M$ with respect to  $I_C$, $II_C$, and $III_C$ by 
\begin{equation}
 \begin{aligned}
\frac{\partial W_{M}}{\partial I_{C}} &=\frac{\partial W_{M}}{\partial J_{1}} -2\frac{\partial W_{M}}{\partial J_{2}} +\frac{\partial W_{M}}{\partial J_{3}},\\
\frac{\partial W_{M}}{\partial II_{C}} &=\frac{\partial W_{M}}{\partial J_{2}} -\frac{\partial W_{M}}{\partial J_{3}},~ 
\frac{\partial W_{M}}{\partial III_{C}} =\frac{\partial W_{M}}{\partial J_{3}}.
\end{aligned}
\end{equation}
In the case of the fourth-order Murnaghan strain-energy function in Eq. \eqref{Murn-4}, we have
\begin{equation}
\begin{aligned}
\frac{\partial W_{M}}{\partial I_{C}} =&( a_{5} -2a_{1}) +( 2a_{2} -2a_{3} +a_{6}) J_{1} +( a_{3} -4a_{8}) J_{2} \\
&+( 3a_{4} -2a_{7}) J_{1}^{2} +4a_{9} J_{1}^{3} +2a_{7} J_{1} J_{2} +a_{6} J_{3}+O\left( \varepsilon^{4}\right),\\
 \frac{\partial W_{M}}{\partial II_{C}}=&( a_{1} -a_{5}) +( a_{3} -a_{6}) J_{1} +a_{7} J_{1}^{2} +2a_{8} J_{2}+O\left( \varepsilon^{3}\right),\\
\frac{\partial W_{M}}{\partial III_{C}}=&a_{5} +a_{6} J_{1}+O\left( \varepsilon^{2}\right).
\end{aligned}
\label{dW/dJ-3}
\end{equation}
Moreover, recalling that $\mathcal{J}\mathrm{=det}(\mathbf{F}) =III_{C}^{1/2}$ and using the Eq. \eqref{III_b-3}, we have
\begin{equation} \begin{aligned}
\frac{2}{\mathcal{J}} =2-e+\frac{3}{4} e^{2} -K-\alpha -\frac{5}{8} e^{3} +\frac{3}{2} eK+\frac{1}{2} e\alpha +\beta -L +O\left( \varepsilon^{4}\right).
\label{2/J-3}
\end{aligned} \end{equation}
Finally, substituting the Eqs. \eqref{I_b}, \eqref{II_b-3}, \eqref{III_b-3}, \eqref{eq-59}, \eqref{dW/dJ-3}, and \eqref{2/J-3} into \eqref{CauchyS} and ignoring the higher-order terms, we can obtain the Cauchy stress up to the third-order terms to be 
\begin{equation}
\begin{aligned} 
\mathbf{t} =\mathbf{t}_1+\mathbf{t}_2+\mathbf{t}_3+O\left( \varepsilon^{4}\right)
\end{aligned} 
\end{equation}
in which the third-order term is 
\begin{equation}
\begin{aligned}
\mathbf{t}_3 =
&\left[\left( -\frac{3}{4} a_{1} -2a_{2} +a_{3} +6a_{4} -2a_{7}\right) e^{2} +(a_{1} +2a_{3} -4a_{8}) K\right.\\
&+\left.( 3a_{1} +4a_{2} -2a_{3} -2a_{5}) \alpha\right] \mathbf{e}+(a_{1} -2a_{3} -a_{5} +2a_{6}) e\mathbf{K} \\
&+( 3a_{1} +4a_{2} -2a_{3} -2a_{5})e\bm{\alpha } +( 4a_{5} -4a_{1})\bm{\beta}\\
& +\left(\left(\frac{3}{4} a_{1} + \frac{3}{2} a_{2} -a_{3} -3a_{4} +2a_{7} +8a_{9}\right)e^{3} 
+( -2a_{1} -2a_{2} +a_{3} +4a_{7} +4a_{8}) eK\right.\\
&\left.+( -2a_{1} -4a_{2} +6a_{3} +12a_{4} +2a_{5}) e\alpha +( 2a_{5} +2a_{6}) L-( 2a_{3} +2a_{5}) \beta \right)\mathbf{I}.
\end{aligned}
\end{equation}

\footnotesize
\begin{center}
\captionof{table}{Transformations of some commonly used scalar invariants of strain tensors.}
\renewcommand{\arraystretch}{2.5}
\hspace*{-0.5cm}
\begin{tabular}{ | m{4cm} | m{4.3cm}| m{3.7cm} |  m{3.75cm}|  } 
  \hline
  & The principal invariants of $\mathbf{C}$ & Landau invariants of $\mathbf{E}$ & 
  Murnaghan invariants of $2\mathbf{E} $ \\ 
  \hline
  The principal invariants of $\mathbf{C}$    \newline
$I_{C} =\mathrm{tr}(\mathbf{C})$    \newline
$II_{C} =\frac{1}{2}\left( \mathrm{tr(}\mathbf{C}\mathrm{)^{2} -tr}\left(\mathbf{C}^{2}\right)\right)$    \newline
$III_{C} =\mathrm{det}\mathbf{C}$
   & 
  \center 
  --
  & 
  $I_{C} =3+2\bar{I}_{1}$\newline
  $II_{C} =3+4\bar{I}_{1} +2\bar{I}_{1}^{2} -2\bar{I}_{2}$\newline
  $ \begin{aligned}
\mathnormal{III_{C} =} 1+2\bar{I}_{1} +2\bar{I}_{1}^{2} -2\bar{I}_{2}\\
+\frac{4}{3}\bar{I}_{1}^{3} -4\bar{I}_{2}\bar{I}_{1} +\frac{8}{3}\bar{I}_{3}
\end{aligned}$\newline
  & 
  $I_{C} =3+J_{1}$\newline
  $II_{C} =3+2J_{1} +J_{2}$\newline   
  $III_{C} =1+J_{1} +J_{2} +J_{3}$
  \\
  \hline
  Landau invariants of $\mathbf{E}$    \newline
$\bar{I}_{1} =\mathrm{tr}(\mathbf{E}) $     \newline
$\bar{I}_{2} =\mathrm{tr}\left(\mathbf{E}^{2}\right) $     \newline
$\bar{I}_{3} =\mathrm{tr}\left(\mathbf{E}^{3}\right)$     \newline
    & 
$\bar{I}_{1} =\frac{1}{2}( -3+I_{C})$      \newline
$\bar{I}_{2} =\frac{1}{4}\left( 3-2I_{C} +I_{C}^{2} -2II_{C}\right)$     \newline
$ \begin{aligned}
\bar{I}_{3} =\frac{1}{8}( 24-24I_{C} +12I_{C}^{2} -2I_{C}^{3}\\
                          -12II_{C} +3I_{C} II_{C} +3III_{C})
\end{aligned}$\newline
    & 
    \center
    --
     & 
    $\bar{I}_{1} =\frac{1}{2} J_{1}$\newline
     $\bar{I}_{2} =\frac{1}{4}\left( J_{1}^{2} -2J_{2}\right)$\newline
     $\bar{I}_{3} =\frac{1}{8}\left( J_{1}^{3} -3J_{1} J_{2} +3J_{3}\right)$\newline 
\\
  \hline
Murnaghan invariants of $2\mathbf{E} $     \newline
$J_{1} =\mathrm{tr}( 2\mathbf{E})  $     \newline
$J_{2} =\frac{1}{2}\left( \mathrm{tr(} 2\mathbf{E}\mathrm{)^{2} -tr}\left( 4\mathbf{E}^{2}\right)\right) $     \newline
$J_{3} =\mathrm{det}( 2\mathbf{E}) $     \newline
    & 
$J_{1} =I_{C} -3$\newline
$J_{2} =3-2I_{C} +II_{C}$\newline
$J_{3}=I_{C}-II_{C}+III_{C}-1$
    & 
 $J_{1} =2\bar{I}_{1}$\newline
 $J_{2} =2\left(\bar{I}_{1}^{2} -\bar{I}_{2}\right)$\newline
 $J_{3} =\frac{4}{3}\left(\bar{I}_{1}^{3} -3\bar{I}_{1}\bar{I}_{2} +2\bar{I}_{3}\right)$\newline
     & 
    \quad\quad\quad\quad\quad\quad\quad --
     \newline
   \\ 
  \hline
\end{tabular}
\label{table-1}
\end{center}

\normalsize

\section{Weekly nonlinear elasticity for isotropic incompressible materials}
Most soft biological materials are assumed to be incompressible, so that $III_{C}=1$, and then the strain-energy function $W$ is a  of just $I_{C}$ and $II_{C}$ and the Cauchy stress tensor is given by
\begin{equation}
 \mathbf{t} =-p\mathbf{I} +2\frac{\partial W}{\partial I_{C}}\mathbf{b} -2\frac{\partial W}{\partial II_{C}}\mathbf{b}^{-1},
 \label{CauchyS-InCom}
\end{equation}
where $p$ is the Lagrange multiplier that needs to be determined by a boundary condition.
Given $III_{C}=1$, then 
\begin{equation}
\begin{aligned}
\bar{I}_{1} =-\bar{I}_{1}^{2} +\bar{I}_{2} +2\bar{I}_{1}\bar{I}_{2} -\frac{2}{3}\bar{I}_{1}^{3} -\frac{4}{3}\bar{I}_{3}, ~\text{and}~
J_{1} = - J_{2} -J_{3}.
\end{aligned}
\label{I_1-J_1}
\end{equation}
In this section, we present the weakly nonlinear expansion of energy functions for commonly used incompressible isotropic hyperelastic solids in the Landau and Murnaghan forms. These include the neo-Hookean model $W_{\text{NH}}$, the two-parameter Mooney-Rivlin model $W_{\text{MR2}}$, and the five-parameter Mooney-Rivlin model $W_{\text{MR5}}$, which are  given by
\begin{equation}
\begin{aligned}
W_{\text{NH}} =&C_{10} (I_{C} -3),\\
W_{\text{MR2}} =&C_{10} (I_{C} -3)+C_{01}(II_{C} -3),\\
W_{\text{MR5}} =&C_{10} (I_{C} -3)+C_{01}(II_{C} -3) +C_{20} (I_{C} -3)^{2} \\
&+C_{11} (I_{C} -3)( II_{C} -3) +C_{02}(II_{C} -3)^{2}.
\end{aligned}
\label{eq-67}
\end{equation}

\subsection{Second-order elasticity}
\subsubsection{strain-energy functions}
For the incompressible materials, the Eq. \eqref{I_1-J_1} indicates that $\bar I_1=\frac{1}{2}(e+\alpha)$ and $J_1=e+\alpha$ are second-order $O\left( \varepsilon^{2}\right)$ quantities.
In addition, recalling that the strain-energy function should be expanded up to the third-order smallness for the second-order elasticity theory, the incompressibility condition in terms of the Landau invariants can be written as 
\begin{equation}
\begin{aligned}
\bar{I}_{1} =\bar{I}_{2} -\frac{4}{3}\bar{I}_{3} + O\left( \varepsilon^{4}\right).
\end{aligned}
\end{equation}
Thus, we have 
\begin{equation}
\begin{aligned}
I_{C} -3&=2\bar{I}_{1} =2\bar{I}_{2} -\frac{8}{3}\bar{I}_{3}  + O\left( \varepsilon^{4}\right),\\
II_{C} -3&=4\bar{I}_{1} +2\bar{I}_{1}^{2} -2\bar{I}_{2} = 2\bar{I}_{2} -\frac{16}{3}\bar{I}_{3}  + O\left( \varepsilon^{4}\right).
\end{aligned}
\end{equation}
Substituting this equation to Eq. \eqref{eq-67}, we can obtain the expansion form of the energy functions as follow:
\begin{equation}
\begin{aligned}
W_{\text{NH}} =&2 C_{10}\bar{I}_{2} -\frac{8}{3} C_{10}\bar{I}_{3} + O\left( \varepsilon^{4}\right),\\
W_{\text{MR2}} =&2( C_{10} +C_{01})\bar{I}_{2} -\frac{8}{3}( C_{10} +2C_{01})\bar{I}_{3} + O\left( \varepsilon^{4}\right),\\
W_{\text{MR5}} =&2( C_{10} +C_{01})\bar{I}_{2} -\frac{8}{3}( C_{10} +2C_{01})\bar{I}_{3} + O\left( \varepsilon^{4}\right).
\end{aligned}
\end{equation}

Following the third-order incompressible isotropic elasticity analysis by \cite{destrade2010third},  the weakly nonlinear expansion of the strain-energy functions  in terms of Landau invariants up to third-order is
\begin{equation}
\begin{aligned}
W_\text{L}=\mu \bar I_2+\frac{\bar A}{3} \bar I_3+O\left( \varepsilon^{4}\right),
\end{aligned}
\label{WL-3-incom}
\end{equation}
From this, we can obtain the following connections between the material constants of the incompressible neo-Hookean,  two-parameter Mooney-Rivlin, and five-parameter Mooney-Rivlin solids 
\begin{equation}
\begin{aligned}
W_\text{NH}: &\quad \mu =2 C_{10} ,\quad \bar{A} =-8 C_{10},\\
W_\text{MR2}: &\quad \mu =2( C_{10} +C_{01}),\quad \bar{A} =-8( C_{10} +2C_{01}),\\
W_\text{MR5}: &\quad \mu =2( C_{10} +C_{01}),\quad \bar{A} =-8( C_{10} +2C_{01}).
\end{aligned}
\end{equation}

Next, recalling the connections between  $I_C, II_C$, $III_C$ and $J_1, J_2$, $J_3$ in Eq. \eqref{I-J}, the  incompressibility condition in Eq. \eqref{I_1-J_1} can be expressed in terms of the Murnaghan invariants as follow:
\begin{equation}
\begin{aligned}
I_{C} -3&=J_{1} = - J_{2} -J_{3},\\
II_{C} -3&=2J_{1} +J_{2} = - J_{2} -2J_{3}.
\end{aligned}
\label{eq-76}
\end{equation}
Substituting \eqref{eq-76} into \eqref{eq-67} and neglecting the terms of $O\left( \varepsilon^{4}\right)$, we can rewrite the stain energy functions as
\begin{equation}
\begin{aligned}
W_{\text{NH}} =&- C_{10} J_{2} - C_{10} J_{3},\\
W_{\text{MR2}} =&-( C_{10} +C_{01}) J_{2} -( C_{10} +2C_{01}) J_{3},\\
W_{\text{MR5}} =&-( C_{10} +C_{01}) J_{2} -( C_{10} +2C_{01}) J_{3} +O\left( \varepsilon^{4}\right).
\end{aligned}
\end{equation}
According to  Eqs. \eqref{WL-3-incom} and \eqref{eq-30}, the strain-energy functions  in terms of Murnaghan invariants becomes
\begin{equation}
\begin{aligned}
W_\text{M}=a_1 J_2+a_5  J_3+O\left( \varepsilon^{4}\right)
\end{aligned}.
\end{equation}
Thus, for the incompressible neo-Hookean,  two-parameter Mooney-Rivlin, and five-parameter Mooney-Rivlin models,  we have the following connections between the material constants: 
\begin{equation}
\begin{aligned}
W_\text{NH}: &\quad a_1=-C_{10}, \quad a_5= -C_{10},\\
W_\text{MR2}: &\quad a_1=-( C_{10} +C_{01}), \quad a_{5} = -( C_{10} +2C_{01}),\\
W_\text{MR5}: &\quad a_1=-( C_{10} +C_{01}), \quad a_{5} = -( C_{10} +2C_{01}).
\end{aligned}
\label{eq-79}
\end{equation}

\subsubsection{Stress-strain relationships}
To simplify the calculation, we use the third-order Murnaghan form strain-energy function to derive the second-order Cauchy stress tensor.
First, using the chain rule, we can rewrite the derivatives of the strain-energy function $W$ with respect to  $I_C$, $II_C$, and $III_C$ as 
\begin{equation}
\begin{aligned}
\frac{\partial W_{\text{M}}}{\partial I_{C}} &=-2\frac{\partial W_{\text{M}}}{\partial J_{2}} +\frac{\partial W_{\text{M}}}{\partial J_{3}} =-2a_{1} +a_{5},\\
\frac{\partial W_{\text{M}}}{\partial II_{C}} &=\frac{\partial W_{\text{M}}}{\partial J_{2}} -\frac{\partial W_{\text{M}}}{\partial J_{3}} =a_{1} -a_{5}.
\end{aligned}
\label{eq-80}
\end{equation}
According to the Eq. \eqref{eq-39}, the incompressibility condition $III_C=1$ gives 
\begin{equation}
e=-\alpha-K+O\left( \varepsilon^{3}\right),
\end{equation}
which indicates that $e$ is the second-order quantity.
Hence, Eq. \eqref{eq-42} can be reduced to
\begin{equation}
\mathbf{b}^{-1} =\mathbf{K} -\mathbf{e} -\bm{\alpha } +( 1-K)\mathbf{I} \ +O\left(\varepsilon^{3}\right).
\label{eq-82}
\end{equation}
Therefore, substituting the Eqs. \eqref{eq-80}, \eqref{eq-82} into \eqref{CauchyS-InCom} and ignoring the higher-order terms, we have the second-order Cauchy stress tensor 
\begin{equation}
\mathbf{t} 
=-2a_{1}\mathbf{e} -( 2a_{1} -2a_{5})\mathbf{K} -2a_{1}\bm{\alpha } +( -6a_{1} +4a_{5} +( 2a_{1} -2a_{5}) K-p)\mathbf{I}+O\left(\varepsilon^{3}\right).
\end{equation}
Specifically, according to the Eq. \eqref{eq-79}, for incompressible neo-Hookean solid, the Cauchy stress tensor is given by
\begin{equation}
\mathbf{t} =2\mathrm{C_{10}}\mathbf{e} +2\mathrm{C_{10}}\bm{\alpha } +( 2C_{10} -p)\mathbf{I}+O\left(\varepsilon^{3}\right).
\end{equation}
For incompressible two-parameter Mooney-Rivlin solid, the Cauchy stress tensor can be rewritten as
\begin{equation}
\mathbf{t} =2\mathrm{( C_{10} +C_{01})}\mathbf{e} +2\mathrm{( C_{10} +C_{01})}\bm{\alpha } -2\mathrm{C}_{\mathrm{01}}\mathbf{K} +( 2C_{10} -2\mathrm{C_{01}}( 1-K) -p)\mathbf{I}+O\left(\varepsilon^{3}\right),
\end{equation}
and for incompressible five-parameter Mooney-Rivlin solid, the Cauchy stress tensor can be expressed by
\begin{equation}
\mathbf{t} =2\mathrm{( C_{10} +C_{01})}\mathbf{e} +2\mathrm{( C_{10} +C_{01})}\bm{\alpha } -2\mathrm{C}_{\mathrm{01}}\mathbf{K} +( 2C_{10} -2\mathrm{C_{01}}( 1-K) -p)\mathbf{I}+O\left(\varepsilon^{3}\right).
\end{equation}

\subsection{Third-order elasticity}
\subsubsection{strain-energy functions}
For the third-order elasticity, the strain-energy function is expanded to fourth-order. The incompressibility condition, in terms of the Landau invariants, is
\begin{equation}
\begin{aligned}
\bar{I}_{1} =\bar{I}_{2} -\frac{4}{3}\bar{I}_{3} +\bar{I}_{2}^{2} + O\left( \varepsilon^{4}\right).
\end{aligned}
\end{equation}
Therefore,  
\begin{equation}
\begin{aligned}
I_{C} -3&=2\bar{I}_{1} =2\bar{I}_{2} -\frac{8}{3}\bar{I}_{3}  +2\bar{I}_{2}^{2}+ O\left( \varepsilon^{4}\right)\\
II_{C} -3&=4\bar{I}_{1} +2\bar{I}_{1}^{2} -2\bar{I}_{2} = 2\bar{I}_{2} -\frac{16}{3}\bar{I}_{3}  +6\bar{I}_{2}^{2}+ O\left( \varepsilon^{4}\right).
\end{aligned}
\end{equation}
Hence, we can rewrite Eq. \eqref{eq-67} as
\begin{equation}
\begin{aligned}
W_{\text{NH}} =&2 C_{10}\bar{I}_{2} -\frac{8}{3} C_{10}\bar{I}_{3} +2 C_{10}\bar{I}_{2}^{2} + O\left( \varepsilon^{4}\right)\\
W_{\text{MR2}} =&2( C_{10} +C_{01})\bar{I}_{2} -\frac{8}{3}( C_{10} +2C_{01})\bar{I}_{3} +2( 3C_{01} +C_{10})\bar{I}_{2}^{2} +O\left( \varepsilon^{4}\right)\\
W_{\text{MR5}} =&2( C_{10} +C_{01})\bar{I}_{2} -\frac{8}{3}( C_{10} +2C_{01})\bar{I}_{3} \\
&+2( 3C_{01} +C_{10} +2C_{20} +2C_{11} +2C_{02})\bar{I}_{2}^{2} +O\left( \varepsilon^{4}\right).
\end{aligned}
\end{equation}

Following \cite{destrade2010third},  the weakly nonlinear expansion of the strain-energy functions  in terms of Landau invariants to  fourth-order is
\begin{equation}
\begin{aligned}
W_\text{L}=\mu \bar I_2+\frac{\bar A}{3} \bar I_3+\bar D \bar I_2^2+O\left(\varepsilon^{5}\right),
\end{aligned}
\label{WL-4-incom}
\end{equation}
where $\bar{D} =\lambda/2+\bar{B}+\Bar{G}$.
Hence, the corresponding material constants of the incompressible neo-Hookean,  two-parameter Mooney-Rivlin, and five-parameter Mooney-Rivlin solids are:
\begin{equation}
\begin{aligned}
W_\text{NH}: &\ \mu =2 C_{10} ,\ \bar{A} =-8 C_{10} ,\ \bar{D} =2 C_{10}\\
W_\text{MR2}: &\  \mu =2( C_{10} +C_{01}) ,\ \bar{A} =-8( C_{10} +2C_{01}) ,\ \ \bar{D} =2( 3C_{01} +C_{10})\\
W_\text{MR5}: &\  \mu =2( C_{10} +C_{01}) ,\ \bar{A} =-8( C_{10} +2C_{01}) ,\\
&\ \bar{D} =2( 3C_{01} +C_{10} +2C_{20} +2C_{11} +2C_{02}).
\end{aligned}
\end{equation}

To represent the strain-energy function in terms of the Murnaghan expansion, we substitute  Eq. \eqref{eq-76} into \eqref{eq-67} to get
\begin{equation}
\begin{aligned}
W_{\text{NH}} =&- C_{10} J_{2} - C_{10} J_{3},\\
W_{\text{MR2}} =&-( C_{10} +C_{01}) J_{2} -( C_{10} +2C_{01}) J_{3},\\
W_{\text{MR5}} =&-( C_{10} +C_{01}) J_{2} -( C_{10} +2C_{01}) J_{3} +( C_{20} +C_{11} +C_{02}) J_{2}^{2}+ O\left( \varepsilon^{4}\right).
\end{aligned}
\end{equation}
Similarly, from \eqref{WL-4-incom} and \eqref{eq-30}, the expansion of the strain-energy functions  in terms of Murnaghan invariants is
\begin{equation}
\begin{aligned}
W_\text{M}=a_1 J_2+a_5  J_3 +b_1 J_2^2+O\left( \varepsilon^{4}\right),
\end{aligned}
\label{WJ-4-incom}
\end{equation}
where $b_1$ is $O\left( \varepsilon\right)$ material constant.
Therefore, for the incompressible neo-Hookean,  two-parameter Mooney-Rivlin, and five-parameter Mooney-Rivlin solids, we have the relationships between the material constants are
\begin{equation}
\begin{aligned}
W_\text{NH}: &~ a_1=-C_{10}, ~ a_5= -C_{10},~  b_1=0,\\
W_\text{MR2}: &~ a_1=-( C_{10} +C_{01}), ~ a_{5} = -( C_{10} +2C_{01}),~  b_1=0,\\
W_\text{MR5}: &~ a_1=-( C_{10} +C_{01}), ~ a_{5} = -( C_{10} +2C_{01}),
~  b_1=( C_{20} +C_{11} +C_{02}).
\end{aligned}
\label{eq-94}
\end{equation}
\subsubsection{Stress-strain relationships in third-order elasticity}
Similarly, we use the fourth-order strain-energy function in Murnaghan expansion to solve the third-order Cauchy stress tensor.
Then, based on Eq. \eqref{WJ-4-incom}, we have 
\begin{equation}
\begin{aligned}
\frac{\partial W_{\text{M}}}{\partial I_{C}} &=-2\frac{\partial W_{\text{M}}}{\partial J_{2}} +\frac{\partial W_{\text{M}}}{\partial J_{3}} =-2a_{1} +a_{5}- 4 b_1 J_2,\\
\frac{\partial W_{\text{M}}}{\partial II_{C}} &=\frac{\partial W_{\text{M}}}{\partial J_{2}} -\frac{\partial W_{\text{M}}}{\partial J_{3}} =a_{1} -a_{5}+2 b_1 J_2.
\end{aligned}
\label{eq-95}
\end{equation}
Referring to the Eq. \eqref{III_b-3}, the incompressibility condition $III_C=1$ gives 
\begin{equation}
e=-\alpha-K+\beta-L+O\left( \varepsilon^{4}\right)
\end{equation}
With this equation, the Eq. \eqref{eq-59} can be rewritten as 
\begin{equation}
\mathbf{b}^{-1} =\mathbf{K} -\mathbf{e} -\bm{\alpha } +( 1-K-L)\mathbf{I} -2\bm \beta+\alpha \mathbf{e} +O\left( \varepsilon^{4}\right).
\label{eq-97}
\end{equation}
In addition, considering the $e$ is $O\left( \varepsilon^{2}\right)$, the $J_2$ in Eq. \eqref{eq-60} can be reduced to 
\begin{equation}
J_{2} =K-\beta +O\left( \varepsilon^{4}\right).
\end{equation}
Therefore, substituting  \eqref{eq-95} and \eqref{eq-97} into \eqref{CauchyS-InCom}, we can derive the Cauchy stress tensor by
\begin{equation}
\begin{aligned}
\mathbf{t} 
= &-2a_{1}\mathbf{e} -( 2a_{1} -2a_{5})\mathbf{K} -2a_{1}\bm{\alpha } -4b_{1}K\mathbf{e} +\mathrm{( 2a_{1} -2a_{5})} \alpha \mathbf{e} -\mathrm{( 4a_{1} -4a_{5})}\bm{\beta }\\
&+( -6a_{1} +4a_{5} +( 2a_{1} -2a_{5} -12b_{1}) K+( 2a_{1} -2a_{5}) L-12b_{1} \beta -p)\mathbf{I}.
\end{aligned}
\end{equation}
Specifically, according to the Eq. \eqref{eq-94}, for incompressible neo-Hookean solid, the Cauchy stress tensor is given by
\begin{equation}
\mathbf{t} =2\mathrm{C_{10}}\mathbf{e} +2\mathrm{C_{10}}\bm{\alpha } +( 2C_{10} -p)\mathbf{I}+O\left( \varepsilon^{4}\right).
\end{equation}
For incompressible two-parameter Mooney-Rivlin solid, the Cauchy stress tensor is given by
\begin{equation}
\begin{aligned}
\mathbf{t} =&2\mathrm{( C_{10} +C_{01})}\mathbf{e} +2\mathrm{( C_{10} +C_{01})}\bm{\alpha } -2\mathrm{C}_{\mathrm{01}}\mathbf{K} + 2C_{01} \alpha \mathbf{e} -4 C_{01}\bm{\beta } \\
&+( 2C_{10} -2\mathrm{C_{01}}( 1-K-L) -p)\mathbf{I}+O\left( \varepsilon^{4}\right).
\end{aligned}
\end{equation}
For incompressible five-parameter Mooney-Rivlin solid, the Cauchy stress tensor is given by
\begin{equation}
\begin{aligned}
\mathbf{t} =&2\mathrm{( C_{10} +C_{01})}\mathbf{e} +2\mathrm{( C_{10} +C_{01})}\bm{\alpha } -2\mathrm{C}_{\mathrm{01}}\mathbf{K}\\
&-4( C_{20} +C_{11} +C_{02})K\mathbf{e}+ 2C_{01} \alpha \mathbf{e} -4 C_{01}\bm{\beta }\\
&+( 2C_{10} -2\mathrm{C_{01}}( 1-K-L) +12( C_{20} +C_{11} +C_{02})( \beta -K) -p)\mathbf{I}+O\left( \varepsilon^{4}\right).
\end{aligned}
\end{equation}

\section{Conclusion}
 nonlinear elastic behaviour of soft materials is of significant importance across many fields, including biology, materials science, geophysics, and acoustics. In this paper, we have given new results for the expansion of the strain-energy functions and Cauchy stress tensor to $O\left( \varepsilon^{4}\right)$ where $\varepsilon=\sqrt{\mathbf{H}\cdot \mathbf{H}}$ and $0<\varepsilon\leq 1$ for the weakly nonlinear asymptotic expansion for small perturbations to the deformation gradient tensor $\mathbf{F} = \mathbf{I} + \mathbf{H}$.
These theories provide us with powerful tools for understanding and predicting the mechanical response of soft materials under complex loading conditions.

By examining distinct invariants of strain tensors, strain-energy functions, stress-strain relations, and transformation relations of material parameters, we reveal the connections between different elastic theories and expand the energy density function to third-order and fourth-order terms under the framework of weak nonlinear theory. Such efforts not only guide further research on the elastic behavior of soft materials but also contribute to finding solutions for practical problems. 
It is worth highlighting that this paper also addresses the strain-energy function and stress-strain relationship of soft materials under incompressible conditions. This consideration facilitates the modelling and analysis of practical problems while providing a simplified approach to tackling complex problems.

\section*{Acknowledgement}
The authors thank the EPSRC for funding this research through grants EP/S030875/1 and EP/S020950/1. Y.D. also acknowledges support from the European Union GA n◦101105740 “Multi-scale and Multi-physics Modelling of Soft Tissues - MULTI-SOFT.” (The views and opinions expressed are those of the author(s) only and do not necessarily reflect those of the European Union or the European Research Executive Agency. Neither the European Union nor the granting authority can be held responsible for them.)

\bibliographystyle{unsrt}  
\bibliography{references}

\begin{thebibliography}{10}

\bibitem{li2012mechanics}
Bo~Li, Yan-Ping Cao, Xi-Qiao Feng, and Huajian Gao.
\newblock Mechanics of morphological instabilities and surface wrinkling in
  soft materials: a review.
\newblock {\em Soft Matter}, 8(21):5728--5745, 2012.

\bibitem{kuhl2003theory}
Ellen Kuhl and Paul Steinmann.
\newblock Theory and numerics of geometrically non-linear open system
  mechanics.
\newblock {\em International Journal for Numerical Methods in Engineering},
  58(11):1593--1615, 2003.

\bibitem{alijani2014non}
Farbod Alijani and Marco Amabili.
\newblock Non-linear vibrations of shells: A literature review from 2003 to
  2013.
\newblock {\em International journal of non-linear mechanics}, 58:233--257,
  2014.

\bibitem{wang2023strain}
Yafei Wang, Yangkun Du, and Fan Xu.
\newblock Strain stiffening retards growth instability in residually stressed
  biological tissues.
\newblock {\em Journal of the Mechanics and Physics of Solids}, page 105360,
  2023.

\bibitem{destrade2023canceling}
M~Destrade, Y~Du, J~Blackwell, N~Colgan, and V~Balbi.
\newblock Canceling the elastic poynting effect with geometry.
\newblock {\em Physical Review E}, 107(5):L053001, 2023.

\bibitem{ogden1997non}
Raymond~W Ogden.
\newblock {\em Non-linear elastic deformations}.
\newblock Courier Corporation, 1997.

\bibitem{rivlin1953solution}
RS~Rivlin.
\newblock The solution of problems in second order elasticity theory.
\newblock In {\em Collected Papers of RS Rivlin: Volume I and II}, pages
  273--301. Springer, 1953.

\bibitem{duan2012effect}
Zheng Duan, Yonghao An, Jiaping Zhang, and Hanqing Jiang.
\newblock The effect of large deformation and material nonlinearity on gel
  indentation.
\newblock {\em Acta Mechanica Sinica}, 28:1058--1067, 2012.

\bibitem{cai1999imperfection}
Zongxi Cai and Yibin Fu.
\newblock On the imperfection sensitivity of a coated elastic half-space.
\newblock {\em Proceedings of the Royal Society of London. Series A:
  Mathematical, Physical and Engineering Sciences}, 455(1989):3285--3309, 1999.

\bibitem{saccomandi2021some}
Giuseppe Saccomandi and Luigi Vergori.
\newblock Some remarks on the weakly nonlinear theory of isotropic elasticity.
\newblock {\em Journal of Elasticity}, 147(1-2):33--58, 2021.

\bibitem{landau1986theory}
Lev~Davidovich Landau, Evgenii~Mikhailovich Lifshitz, Arnold~Markovich
  Kosevich, and Lev~Petrovich Pitaevskii.
\newblock {\em Theory of elasticity: volume 7}, volume~7.
\newblock Elsevier, 1986.

\bibitem{murnaghan1937finite}
Francis~Dominic Murnaghan.
\newblock Finite deformations of an elastic solid.
\newblock {\em American Journal of Mathematics}, 59(2):235--260, 1937.

\bibitem{ogden2004fitting}
Raymond~W Ogden, Giuseppe Saccomandi, and Ivonne Sgura.
\newblock Fitting hyperelastic models to experimental data.
\newblock {\em Computational Mechanics}, 34:484--502, 2004.

\bibitem{destrade2002incompressible}
Michel Destrade, Paul~A Martin, and Tom~CT Ting.
\newblock The incompressible limit in linear anisotropic elasticity, with
  applications to surface waves and elastostatics.
\newblock {\em Journal of the Mechanics and Physics of Solids},
  50(7):1453--1468, 2002.

\bibitem{destrade2010onset}
Michel Destrade, Michael~D Gilchrist, and Jerry~G Murphy.
\newblock Onset of nonlinearity in the elastic bending of blocks.
\newblock 2010.

\bibitem{krishna2012interaction}
Vamshi Krishna~Chillara and Cliff~J Lissenden.
\newblock Interaction of guided wave modes in isotropic weakly nonlinear
  elastic plates: Higher harmonic generation.
\newblock {\em Journal of Applied Physics}, 111(12), 2012.

\bibitem{sabin1983contact}
GCW Sabin and PN~Kaloni.
\newblock Contact problem of a rigid indentor in second order elasticity
  theory.
\newblock {\em Zeitschrift f{\"u}r angewandte Mathematik und Physik ZAMP},
  34(3):370--386, 1983.

\bibitem{du2023nonlinear}
Yangkun Du, Peter Stewart, Nicholas~A Hill, Huabing Yin, Raimondo Penta, Jakub
  K{\"o}ry, Xiaoyu Luo, and Raymond Ogden.
\newblock Nonlinear indentation of second-order hyperelastic materials.
\newblock {\em Journal of the Mechanics and Physics of Solids}, 171:105139,
  2023.

\bibitem{destrade2010third}
Michel Destrade and Raymond~W Ogden.
\newblock On the third-and fourth-order constants of incompressible isotropic
  elasticity.
\newblock {\em The Journal of the Acoustical Society of America},
  128(6):3334--3343, 2010.

\end{thebibliography}

\end{document}